\documentclass[twocolumn,epjc3]{svjour3}          

\RequirePackage[T1]{fontenc}

\smartqed  

\RequirePackage{graphicx}
\RequirePackage{mathptmx}      
\RequirePackage{flushend}
\RequirePackage[numbers,sort&compress]{natbib}
\RequirePackage[colorlinks,citecolor=blue,urlcolor=blue,linkcolor=blue]{hyperref}

\def\beq{\begin{equation}}
\def\eeq{\end{equation}}
\def\bear{\begin{eqnarray}}
\def\ear{\end{eqnarray}}

\usepackage{enumitem}
\usepackage{graphicx,subfigure}
\usepackage{dcolumn}
\usepackage{bm}
\usepackage{color}

\journalname{Eur. Phys. J. C}

\begin{document}

\title{Energy conditions of non-singular black hole spacetimes in conformal gravity}

\author{Bobir Toshmatov\thanksref{e1,1,2} \and Cosimo Bambi\thanksref{e2,3,4} \and Bobomurat Ahmedov\thanksref{e3,2,5} \and Ahmadjon Abdujabbarov\thanksref{e4,2,5,6} \and Zden\v{e}k Stuchl\'{i}k\thanksref{e5,1}
}

\thankstext{e1}{e-mail: bobir.toshmatov@fpf.slu.cz}
\thankstext{e2}{e-mail: bambi@fudan.edu.cn}
\thankstext{e3}{e-mail: ahmedov@astrin.uz}
\thankstext{e4}{e-mail: ahmadjon@astrin.uz}
\thankstext{e5}{e-mail: zdenek.stuchlik@fpf.slu.cz}

\institute{Faculty of Philosophy and Science, Institute of Physics, Silesian University in Opava, Bezru\v{c}ovo n\'{a}m\v{e}st\'{i} 13, CZ-74601 Opava, Czech Republic\label{1} \and
Ulugh Beg Astronomical Institute, Astronomicheskaya 33, Tashkent 100052, Uzbekistan\label{2}\and
Center for Field Theory and Particle Physics and Department of Physics, Fudan University, 200433 Shanghai, China\label{3}\and
Theoretical Astrophysics, Eberhard-Karls Universit\"{a}t T\"{u}bingen, 72076 T\"{u}bingen, Germany\label{4} \and
National University of Uzbekistan, Tashkent 100174, Uzbekistan\label{5} \and
Tashkent University of Information Technologies, 108 Amir Temur Avenue, Tashkent 100200, Uzbekistan\label{6}}

\date{Received: date / Accepted: date}

\maketitle

\begin{abstract}
Conformal gravity can elegantly solve the problem of spacetime singularities present in Einstein's gravity. For every physical spacetime, there is an infinite family of conformally-equivalent singularity-free metrics. In the unbroken phase, every non-singular metric is equivalent and can be used to infer the physical properties of the spacetime. In the broken phase, a Higgs-like mechanism should select a certain vacuum, which thus becomes the physical one. However, in the absence of the complete theoretical framework we do not know how to select the right vacuum. In this paper, we study the energy conditions of non-singular black hole spacetimes obtained in conformal gravity assuming they are solutions of Einstein's gravity with an effective energy-momentum tensor. We check whether such conditions can be helpful to select the vacuum of the broken phase.
\end{abstract}

\section{Introduction}\label{sec-intr}

The presence of spacetime singularities in some physically relevant solutions in Einstein's gravity is an outstanding and longstanding problem. At the singularity, predictability is lost and standard physics breaks down. The resolution of spacetime singularities in Einstein's gravity is often attributed to unknown quantum gravity effects. The topic has been an active research filed from the past decades and different authors have explored different scenarios. Among the many proposals, an appealing and simple solution is represented by the family of conformal theories of gravity~\cite{cg1,cg2,cg3,cg4,cg5}.

In conformal gravity, the theory is invariant under a conformal transformation of the metric tensor
\bear
g_{\mu\nu} \rightarrow g_{\mu\nu}^* = \Omega^2 g_{\mu\nu} \, ,
\ear
where $\Omega = \Omega (x)$ is a function of the point of the spacetime. Examples of conformal theories of gravity in four dimensions are
\bear\label{eq-m1}
\mathcal{L}_1 &=& \phi^2 R + 6 \, g^{\mu\nu} (\partial_\mu \phi)( \partial_\nu \phi) \, , \\
\label{eq-m2}
\mathcal{L}_2 &=& a \, C^{\mu\nu\rho\sigma} C_{\mu\nu\rho\sigma}
+ b \, \tilde{R}^{\mu\nu\rho\sigma} R_{\mu\nu\rho\sigma} \, .
\ear
In the theory in~(\ref{eq-m1}), Einstein's gravity is modified by introducing the auxiliary scalar field $\phi$ (dilaton). In the theory in~(\ref{eq-m2}), $C^{\mu\nu\rho\sigma}$ is the Weyl tensor, $R^{\mu\nu\rho\sigma}$ is the Riemann tensor, $\tilde{R}^{\mu\nu\rho\sigma}$ is the dual of the Riemann tensor, $a$ and $b$ are constants, and there is no dilaton.

Conformal gravity can solve the problem of spacetime singularities present in Einstein's gravity as follows (see Ref.~\cite{Bambi:1611.00865} for more details). If a metric $g_{\mu\nu}$ is singular in a gauge, we can always remove its singularities by performing a suitable conformal transformation. Indeed, if a gauge variant quantity (like a metric tensor) is singular in a gauge, but not in another gauge, its singularities are not physical, but only an artifact of the gauge. Let us note that there is no fine tuning here, because there is always an infinite class of rescaling factors suitable to remove these singularities. Since the theory is invariant under conformal transformations, we have the freedom to choose the gauge, and the physics will be independent of our choice.

In Einstein's gravity, this is not possible because the theory is not invariant under conformal transformations. Moreover, the world around us is not conformally invariant. If conformal invariance is a fundamental symmetry of the Universe, it must be broken. In such a case, it is necessary a Higgs-like mechanism that breaks the symmetry and selects a certain vacuum. The selection of the vacuum is determined by the details of the theory and by how conformal symmetry is broken. The selected vacuum should be one in which the metric is free from singularities, but we do not have currently any selection criterion to decide which vacuum should be chosen among an infinite number of options.

In Refs.~\cite{Bambi:1611.00865,Modesto:2016max}, the authors found singularity-free black hole solutions in conformal gravity.
As proved in Refs.~\cite{Bambi:1611.00865,Modesto:2016max}, these spacetimes are geodetically complete because no massless or massive particles can reach the center of the black hole in a finite amount of time or for a finite value of the affine parameter. Moreover, curvature invariants do not diverge at $r=0$.
While conformal invariance is broken at the quantum level for the theories in~(\ref{eq-m1}) and in~(\ref{eq-m2})~\cite{Fradkin:1981iu}, it is preserved in any finite (i.e. without divergences) quantum field theory of gravity~\cite{Modesto:2014lga}. The latter look thus the natural theoretical framework for conformal symmetry.

The aim of the present paper is to study the ``energy conditions'' of the singularity-free black hole solutions found in~\cite{Bambi:1611.00865,Modesto:2016max}. For energy conditions here we mean the energy conditions of the ``effective'' energy-momentum tensor of these solutions assuming the standard Einstein's equations. For the theory in~(\ref{eq-m1}), this corresponds to the energy-momentum tensor of the dilaton field $\phi$. For the theory in~(\ref{eq-m2}), the actual energy-momentum tensor vanishes, because the singularity-free black hole metrics are vacuum solutions, so we consider an effective energy-momentum tensor. These energy conditions may be used as a selection criterion to choose the right vacuum in the broken phase by requiring the minimum possible violation.
Note that this is the most general approach that we can employ if we do not want to limit our discussion to a particular model. As we have already pointed out, our black hole metrics are solutions in an infinite class of conformal gravity theories. If we considered a specific conformal theory of gravity, we could write the field equations of the theory and separate the non-Einstein modes~\cite{maldacena} to better explore the black holes in the theory under consideration.

The content of the paper is as follows. In Section~\ref{sec-spacetime}, we briefly review the singularity-free black hole metrics found in Refs.~\cite{Bambi:1611.00865,Modesto:2016max}. In Section~\ref{energy conditions}, we study the null, the weak, and the strong energy conditions in these black hole spacetimes. Summary and conclusions are reported in Section~\ref{conclusion}. Throughout the paper, we employ natural units in which $G_{\rm N} = c = 1$ and a metric with signature $(-+++)$.

\section{Non-singular black hole spacetimes in conformal gravity}\label{sec-spacetime}

The line element of the singularity-free non-rotating black hole spacetime in conformal gravity found in~\cite{Bambi:1611.00865,Modesto:2016max} can be written as
\bear\label{metric-Schw}
ds^{\ast2}=S(r)ds_{Schw}^2,
\ear
where $ds_{Schw}$ is the line element of the Schwarzschild spacetime in Schwarzschild coordinates
\bear\label{Schwarzschild}
ds_{Schw}^2 &=& -f(r)dt^2+ \frac{dr^2}{f(r)}+r^2d\Omega^2 \, , \nonumber\\
f(r) &=& 1-\frac{2M}{r} \, ,
\ear
and $S(r)$ is the scaling factor
\bear\label{conf1}
S(r)=\left(1+\frac{L^2}{r^2}\right)^{2N} \, ,
\ear
where $N=1$, 2, 3, ... and $L$ is a new scale whose value is not specified by the theory. It is natural to expect $L$ to be either of the order of the Planck scale or of the order of $M$, since there are no other scales in the system.

The line element of the rotating black hole solution is given by~\cite{Bambi:1611.00865}
\bear\label{metric-Kerr}
ds^{\ast2}=S(r,\theta)ds_{Kerr}^2,
\ear
where $ds_{Kerr}$ is the line element of the Kerr spacetime in Boyer-Lindquist coordinates
\bear\label{Kerr-spacetime}
ds_{Kerr}^2 &=&-\left(1-\frac{2Mr}{\Sigma}\right)dt^2- 2\frac{2Mar}{\Sigma}\sin^2\theta dt d\phi +\frac{\Sigma}{\Delta}dr^2\nonumber\\
&&+\Sigma d\theta^2+\left(r^2+a^2+\frac{2Ma^2r}{\Sigma}\sin^2\theta\right)\sin^2\theta d\phi^2,\nonumber\\
\ear
$\Delta=r^2-2Mr+a^2$, $\Sigma=r^2+a^2\cos^2\theta$, $a=J/M$ is the specific angular momentum and $S(r,\theta)$ is the scaling factor
\bear\label{conf2}
S(r,\theta)=\left(1+\frac{L^2}{\Sigma}\right)^{2(N+1)} \, ,
\ear
with $N=1$, 2, 3, ... If we assume that $L \propto M$ and $N = 1$, then astrophysical observations require $L/M < 1.2$~\cite{Bambi:1701.00226}.

The difference between the exponent in~(\ref{conf1}) and in (\ref{conf2}) is due to the fact that, in the non-rotating case, the metric is already non-singular when the exponent is 2, while this is not enough in the rotating case and we need at least that the exponent is 4.

Last, it is worth noting that these solutions are not only stationary exact solutions of conformal gravity theories like those in~(\ref{eq-m1}) and in~(\ref{eq-m2}). As in Einstein's gravity, these black hole solutions can be created from gravitational collapse~\cite{Bambi:2016yne}.

\section{Energy conditions}\label{energy conditions}

In this section we want to study the energy conditions of the non-singular black hole spacetimes in conformal gravity of Ref.~\cite{Bambi:1611.00865}. First, we turn to the diagonal stress-energy tensor $T^{(a)(b)}=diag(\rho, P_1, P_2, P_3)$ by using the following orthonormal tetrads, which correspond to the standard locally non-rotating frame (LNRF)~\cite{BardeenApJ:1972}:
\begin{eqnarray} \label{tetrad}
e_\mu^{(a)}=\left(
\begin{array}{c c c c}
\sqrt{\mp(g_{tt}-\Omega g_{t\phi})} & 0 & 0 & 0 \\
0 & \sqrt{\pm g_{rr}} & 0 & 0 \\
0 & 0 & \sqrt{g_{\theta\theta}} & 0 \\
\Omega\sqrt{g_{t\phi}} & 0 & 0 & \sqrt{g_{\phi\phi}} \\
\end{array} \right).
\end{eqnarray}
where $\Omega=g_{t\phi}/g_{\phi\phi}$ is the angular velocity of the black hole. The energy-momentum tensor is turned to diagonal form by the relation
\bear\label{contra}
T^{(a)(b)}=\frac{1}{8\pi}e_\mu^{(a)}e_\nu^{(b)}G^{\mu\nu}.
\ear
The signature of (\ref{tetrad}) reflects the considered region. We claim that the spacetime metric~(\ref{metric-Kerr}) has two horizons: inner (Cauchy) and outer (event) horizons, in the case of black hole. For the regions outside the event horizon and inside the inner horizon, we choose the signature of the components of the orthonormal tetrads $e_0^{(0)}$ and $e_1^{(1)}$ as $(-,+)$, respectively. For the region between these two horizons, we take the signature of $e_0^{(0)}$ and $e_1^{(1)}$ as $(+,-)$, respectively.

\subsection{Null Energy Condition}

The null energy condition (NEC) can be written in the following compact form
\bear
T_{\mu\nu}n^\mu n^\nu\geq0\ ,
\ear
where $n^\mu$ is a generic null-like vector. The physical interpretation of the NEC is that the energy density measured by an observer traversing a null curve is never negative. In the LNRF, it corresponds to the condition
\bear\label{nec}
\rho+P_i\geq0, \qquad i=1,2,3.
\ear

\subsubsection{NEC for non-rotating black holes in conformal gravity}\label{nec-schwarzschild}

For the non-rotating black hole metric in~(\ref{metric-Schw}), Eq.~(\ref{nec}) becomes
\bear\label{nec-schw1}
\rho+P_1&=&-\frac{4NL^2r^{4N-3}}{(r^2+L^2)^{2N+2}}\nonumber\\ &&\times\{r^3 + L^2[4M(-1+2N) + (3-4N) r]\},\nonumber\\
\rho+P_2&=&\frac{4NL^2r^{4N-3}}{(r^2+L^2)^{2N+2}}\\
&&\times\{3(3M-r)r^2+L^2(M+4MN+r-2Nr)\}\nonumber\\
&&=\rho+P_3.\nonumber
\ear

\begin{figure*}[th]
\begin{center}
\includegraphics[width=0.44\linewidth]{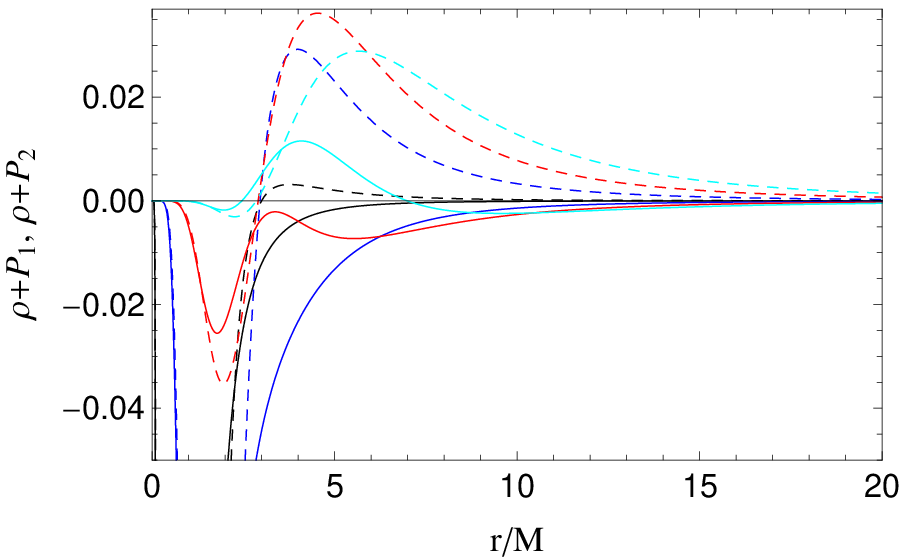}
\includegraphics[width=0.44\linewidth]{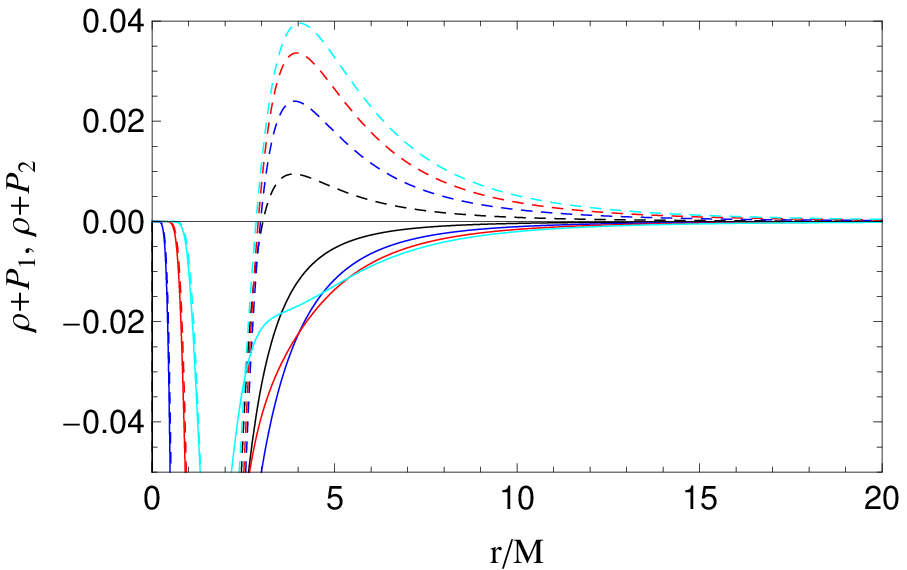}
\end{center}
\caption{\label{fig-nec-schw} Violation of the NEC ($\rho+P_1$ -- solid curves, $\rho+P_2$ -- dashed curves) of the non-singular non-rotating black hole spacetime. \textbf{Left panel:} $N=3$, and the values of the dimensionless parameter $L/M$: $L/M=0.3$ (black curves), $L/M=1.2$ (blue curves), $L/M=2$ (red curves), $L/M=3$ (cyan curves). \textbf{Right panel:} $L/M=1$, and the values of the parameter $N$: $N=1$ (black curves), $N=3$ (blue curves), $N=5$ (red curves), $N=7$ (cyan curves).}
\end{figure*}

Let us note that at the center of the spacetime, $r=0$, the two relations in~(\ref{nec-schw1}) vanish, i.e., the NEC is satisfied, while the NEC~(\ref{nec}) may be violated at larger radii as shown in Fig.~\ref{fig-nec-schw}.

At large distances, $r\rightarrow\infty$, both $\rho+P_1$ and $\rho+P_2$ tend to zero. However, they tend to zero from different directions, i.e.,
\bear\label{assymp}
&&\lim_{r\rightarrow\infty}\left(\rho+P_1\right)=-0,\nonumber\\
&&\lim_{r\rightarrow\infty}\left(\rho+P_2\right)=+0.
\ear
Now we investigate the domains in which the NEC holds and is violated in the regular non-rotating black hole spacetime in conformal gravity. From~(\ref{nec-schw1}) one can easily check that $\rho+P_2$ has always two zeros ($r_1=0$ and $r_2$) irrespective of the values of the parameters $L/M$ and $N$: $\rho+P_2<0$ for $r\in(0,r_2)$ and $\rho+P_2\geq0$ for $r\in[r_2,+\infty)$. On the other hand, $\rho+P_1$ has one zero ($r'_1=0$)~\footnote{In order to distinguish the zeros of $\rho+P_2$ and $\rho+P_1$, we denote the zeros of $\rho+P_1$ with prime ``$'$".}, three zeros ($r'_1=0$, $r'_2$, and $r'_3$), or two zeros ($r'_1=0$, $r'_2=r'_3=r''_2$) depending on the values of the parameters $L/M$ and $N$: $\rho+P_1<0$ for $r\in(0,r'_2)\cup(r'_3,+\infty)$ and $\rho+P_1\geq0$ for $r\in[r'_2,r'_3]$. Let us note that $\rho+P_1$ has three zeros, i.e., the NEC is satisfied in some region of the spacetime with a finite value of $r$, if the parameter $L$ satisfies the following condition:
%
\bear\label{L-crit}
L>L_{cr}\equiv\frac{6\sqrt{3}(2N-1)}{\sqrt{(4N-3)^3}}M \, .
\ear
For $L=L_{cr}$, we have $r'_2 = r'_3$, which we can rename $r''_2$ and is given by
\bear
r''_2=\frac{L^{2/3}(4N-3)+3[2M(2N-1)]^{2/3}}{3[2M(2N-1)]^{1/3}}L^{2/3}.
\ear
\begin{figure}[h]
\centering
\includegraphics[width=0.43\textwidth]{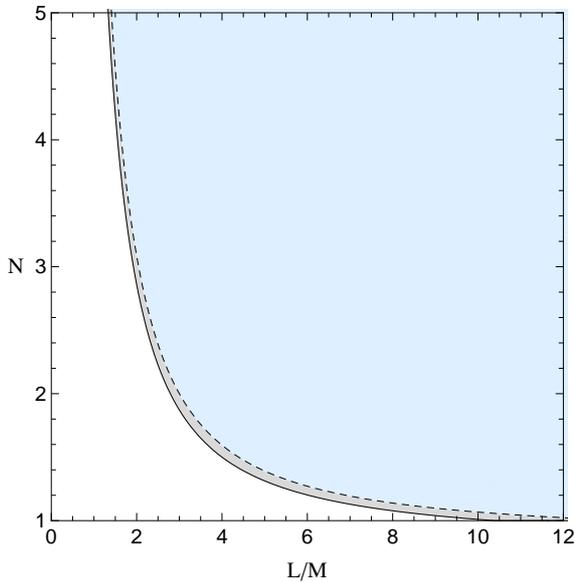}
\caption{\label{fig-nec-schw2} Plot showing the parametric region for the NEC in the non-singular non-rotating black hole spacetime in conformal gravity. In all shaded regions the NEC is satisfied in some particular part of the spacetime. In the light gray region $r'_2>r_2$, while in the light blue region $r'_2<r_2$. The boundary between the white and the light gray regions (black solid curve) corresponds to $L=L_{cr}$ in Eq.~(\ref{L-crit}), while the boundary between the light gray and the light blue regions (black dashed curve) represent the case of $r_2=r'_2$.}
\end{figure}
For $L<L_{cr}$, $\rho+P_1$ has only one zero at $r'_1=0$ and in this case the NEC is violated everywhere in the spacetime except at $r=0$, since, according to~(\ref{nec}), the NEC is satisfied when both $\rho+P_1$ and $\rho+P_2$ are non-negative at the same time. Thus, the NEC is satisfied in the region $r\in\{0\}\cup[r'_2,r'_3]\cap[r_2,r'_3]$, or to be more precise, $r\in\{0\}\cup[r'_2,r'_3]$ if $r'_2>r_2$, gray region in Fig.~\ref{fig-nec-schw2}, or $r\in\{0\}\cup[r_2,r'_3]$ if $r_2>r'_2$, light blue region in Fig.~\ref{fig-nec-schw2}. Thus, the upper bound of the NEC is defined by the largest zero of $\rho+P_1$ ($r'_3$).

The depth of the violation of the NEC in the regular non-rotating black hole spacetime in conformal gravity decreases as the values of the parameters $L$ or $N$ increase. More specifically, if we increase the values of $L$ and $N$, the depth of the violation of the NEC and the region of the violation in terms of the radius $r$ decrease.

\begin{figure}[h]
\centering
\includegraphics[width=0.43\textwidth]{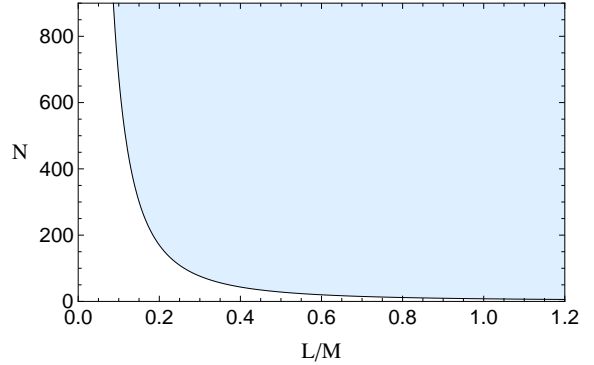}
\caption{\label{fig-nec-schw3} Plot showing the parametric region for the NEC to be satisfied in the region $r\in[r'_2,r'_3]$ (shaded region) when $L/M<1.2$. The border -- black, solid curve, corresponds to the case $r'_2=r'_3$.}
\end{figure}

If we assume that $L \propto M$, astrophysical observations require $L/M<1.2$~\cite{Bambi:1701.00226}. In such a case, as shown in Fig.~\ref{fig-nec-schw}, the NEC is never satisfied except for $r\in\{0\}$, at least for small $N$ ($N<6$). Within such scenario, if we want that the NEC is satisfied in $r\in[r'_2,r'_3]$, the value of $N$ must be large -- see Fig.~\ref{fig-nec-schw3}. If we consider the alternative scenario in which $L$ is of the order of the Planck scale, for $M \gg L$ (astrophysical black holes) the spacetime is very similar to the standard Kerr background in Einstein's gravity, while $L/M = O (1)$ may be possible in the case of Planck-scale black holes.

One can see from Fig.~\ref{fig-nec-schw2} that for large values of the parameter $L$, $r_2>r'_2$, i.e., the inner boundary of the NEC satisfied region is restricted by $r_2$. Interestingly, for large values of the parameter $L$, $r_2$ tends to a finite value
%
%
%
\bear
r_2=\frac{(4N+1)M}{2N-1}+\mathcal{O}\left(\frac{1}{L^2}\right) \, ,
\ear
while the outer boundary of the NEC satisfied region, $r'_3$, monotonically increases with $L$ as
\bear
r'_3=L\sqrt{4N-3}-\frac{2\sqrt{3}(2N-1)}{4N-3}M+\mathcal{O}\left(\frac{1}{L^2}\right).
\ear
In the case of large values of $N$, the NEC is satisfied only outside the event horizon of the black hole, i.e.,
\bear\label{nec-schw-N}
r_2=2M+\mathcal{O}\left(\frac{1}{N}\right),
\ear
while the outer boundary tends to the value
\bear\label{nec-schw-N}
r'_3=\frac{(6N-1)L^2-3M^2}{2\sqrt{3N}L}- M+\mathcal{O}\left(\frac{1}{N}\right).
\ear
Finally, we conclude the subsection with the claim that the NEC is never satisfied inside the event horizon of the non-singular non-rotating black hole spacetime in conformal gravity, except at the center of the spacetime, $r=0$. Increasing the values of the parameters $L/M$ and $N$, the NEC satisfied region increases, tends to infinity starting from the event horizon of the black hole.

\subsubsection{NEC for rotating black holes in conformal gravity}\label{subsec-nec-kerr}

In the case of the non-singular rotating black hole spacetime in conformal gravity, the expressions are more complicated, including the energy-momentum tensor. Therefore, we do not report the full expressions of the components of the energy-momentum tensor in this case because of their cumbersome forms. Full expressions of the energy-momentum tensor in the polar caps ($\theta=0, \pi$) are given in Appendix~\ref{A1}.

Now we start studying the NEC by checking the behavior of the energy-momentum tensor near the origin, i.e. inside the inner horizon. The NEC~(\ref{nec}) in the non-singular rotating black hole spacetime in conformal gravity~(\ref{Kerr-spacetime}) near the center of the spacetime takes the form
\bear\label{nec-kerr-1}
\lim_{r\rightarrow0}\left(\rho+P_1\right) &=&-\frac{4a^2(N+1)(a^2x^2)^{2N}L^2}{(a^2x^2+L^2)^{2N+4}}\{3a^2x^2(x^2-2) \nonumber\\&&+L^2[x^2-4+2N(x^2-1)]\}\ ,\nonumber\\
\lim_{r\rightarrow0}\left(\rho+P_2\right) &=&\frac{4a^2(N+1)(a^2x^2)^{2N}L^2}{(a^2x^2+L^2)^{2N+4}}\{a^2x^4\nonumber\\ &&-L^2[x^2-2+4N(x^2-1)]\},\\
\lim_{r\rightarrow0}\left(\rho+P_3\right) &=&-\frac{4a^2(N+1)(a^2x^2)^{2N}L^2}{(a^2x^2+L^2)^{2N+4}}\{a^2x^2(3x^2-4)\nonumber\\ &&-L^2[x^2-2+4N(x^2-1)]\}\ ,\nonumber
\ear
where $x=\cos\theta$. At the poles of the spacetime, i.e., $x^2=1$, the above relations take the form
\bear\label{nec-kerr-2}
\lim_{r\rightarrow0}\left(\rho+P_1\right)&&=\frac{12a^{4N+2}(N+1)L^2}{(a^2+L^2)^{2N+3}}\ ,\nonumber\\
\lim_{r\rightarrow0}\left(\rho+P_2\right) &&=\frac{4a^{4N+2}(N+1)L^2}{(a^2+L^2)^{2N+3}}\\ &&=\lim_{r\rightarrow0}\left(\rho+P_3\right)\ ,\nonumber
\ear
Eqs.~(\ref{nec-kerr-2}) show that the NEC in the non-singular rotating black hole spacetime is satisfied (at least) at the center of the spacetime. Now we want to study the regions where the NEC is satisfied. As in the non-rotating case, for the rotating one we have that at large radii the NEC relations tend to zero from different directions as~(\ref{assymp}). In this case, $\rho+P_2$ can have two ($r_1$, $r_2$), or one ($r_1=r_2$) positive zeros, or does not have any zero, i.e., it is positive everywhere in the spacetime. If it has two zeros: $\rho+P_2<0$ for $r\in(r_1,r_2)$ and $\rho+P_2\geq0$ for $r\in[0,r_1]\cup[r_2,+\infty)$. If it has one zero or does not have any zero: $\rho+P_2\geq0$ for $r\in[0,+\infty)$. These cases correspond to the rotating no-horizon spacetime, $a\geq M$\footnote{For $L=0$ (Kerr metric), it is well-known that there is no horizon and we have a naked singularity for $a>M$. For a non-vanishing $L$, there is no horizon too when $a>M$ (because the existence of the horizon is independent of the value of $L$), but even the singularity disappears.}.

$\rho+P_1$ can have one ($r'_3$), three ($r'_1$, $r'_2$, $r'_3$), or two ($r'_1=r'_2$, $r'_3$) positive zeros. If it has one or two zeros: $\rho+P_1<0$ for $r\in(r'_3,+\infty)$, and $\rho+P_1\geq0$ for $r\in[0,r'_3]$. If it has three zeros: $\rho+P_2<0$ for $r\in(r'_1,r'_2)$, and $\rho+P_2\geq0$ for $r\in[0,r'_1]\cup[r'_2,r'_3]$. Thus, the NEC is satisfied in the region $r\in[0,r_1]\cap[0,r'_1]\cup[r_2,r'_3]\cap[r'_2,r'_3]$, or, more precisely, in $r\in[0,r_1]\cup[r_2,r'_3]$ if $r_1<r'_1$ and $r_2>r'_2$, $r\in[r_0,r'_1]\cup[r_2,r'_3]$ if $r_1>r'_1$ and $r_2>r'_2$. Moreover, there are other possible scenarios: $r\in[0,r'_1]\cup[r'_2,r'_3]$ if $r_1>r'_1$ and $r_2<r'_2$, $r\in[0,r_1]\cup[r'_2,r'_3]$ if $r_1<r'_1$ and $r_2<r'_2$ which do not exist in our case.

\begin{figure*}[th]
\begin{center}
\includegraphics[width=0.32\linewidth]{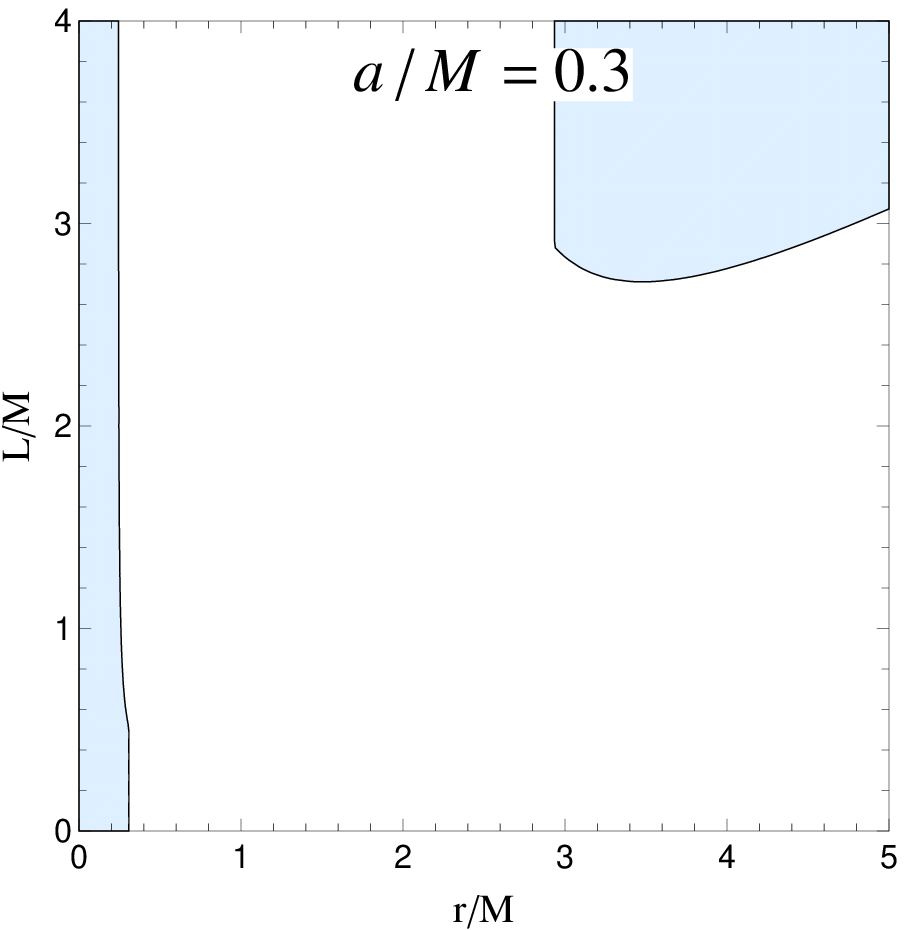}
\includegraphics[width=0.32\linewidth]{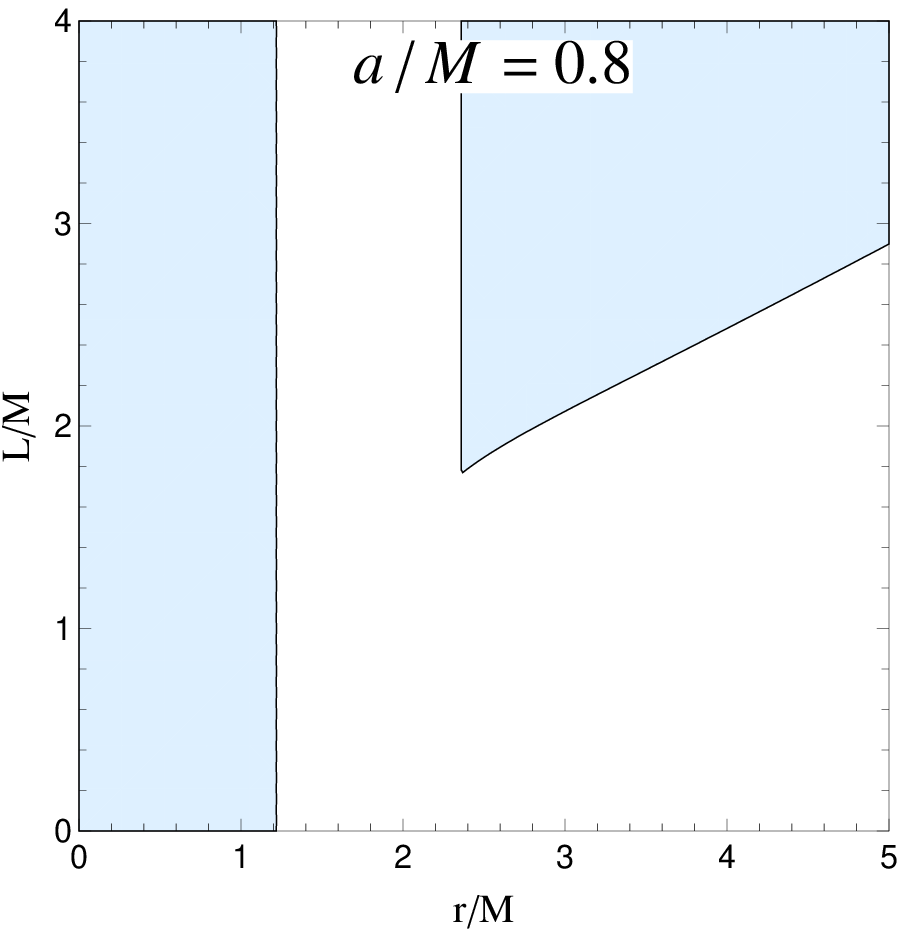}
\includegraphics[width=0.32\linewidth]{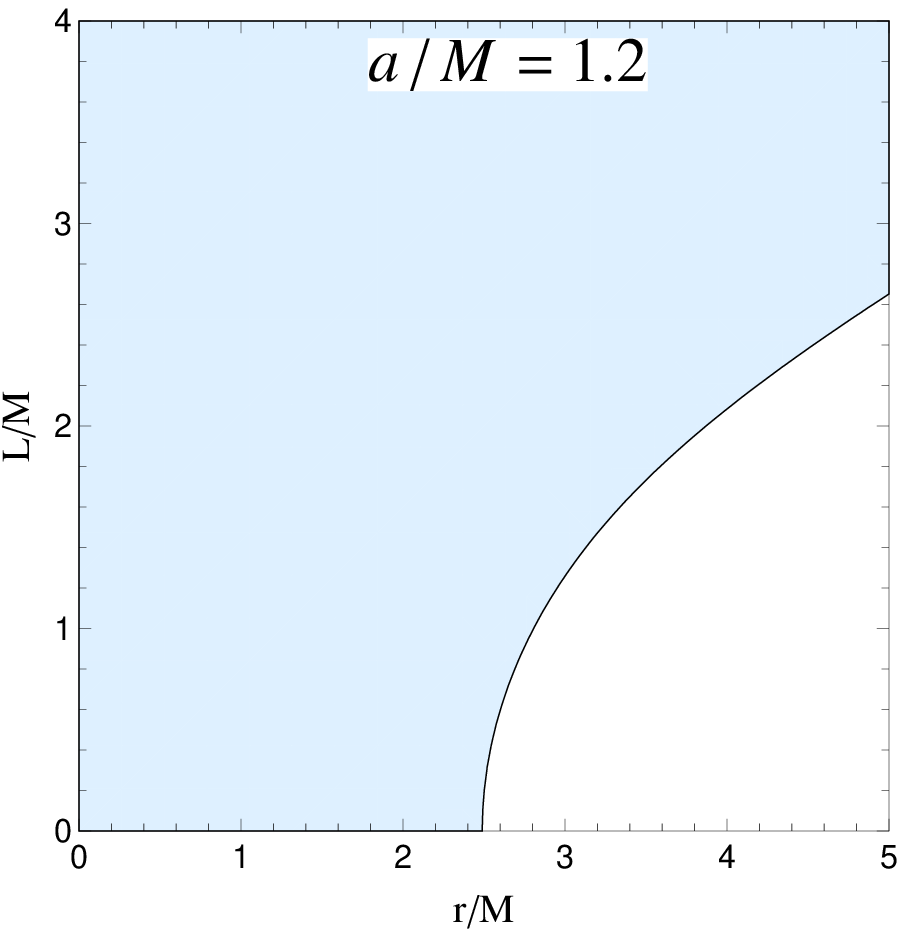}
\end{center}
\caption{\label{fig-nec-kerr1} Plot showing the regions where the NEC is satisfied (shaded regions) at the poles of the non-singular rotating black hole spacetime in conformal gravity. Here $N=1$. }
\end{figure*}

With this result we know that the NEC is satisfied in two regions: inner $r\in[0,r_1]\cap[0,r'_1]$ and outer $r\in[r_2,r'_3]\cap[r'_2,r'_3]$ branches (regions) in the non-singular rotating black hole spacetime in conformal gravity -- see Fig.~\ref{fig-nec-kerr1}. The outer boundary of the inner region of the NEC is defined by the radius $r_1$. For small values of the rotation parameter $a$, $r_1$ is small and close to zero. However, increasing the value of the rotation parameter $a$, it monotonically increases and never vanishes. In the absence of a horizon when $a\geq M$, and large values of $N$, the inner and outer branches of the NEC join, i.e., $r_1$ and $r_2$ vanish and the outer boundary of the NEC is defined by $r'_3$, i.e., $r\in[0,r'_3]$ -- see the right panel of Fig.~\ref{fig-nec-kerr1}. Note that the requirement $L/M<1.2$ (valid for $N=1$ but it can be easily extended to higher values of $N$ obtaining a similar constraint) coming from X-ray data of astrophysical black holes~\cite{Bambi:1701.00226}, corresponds to the inner region of the NEC in Fig.~\ref{fig-nec-kerr1}.

\begin{figure*}[th]
\begin{center}
\includegraphics[width=0.44\linewidth]{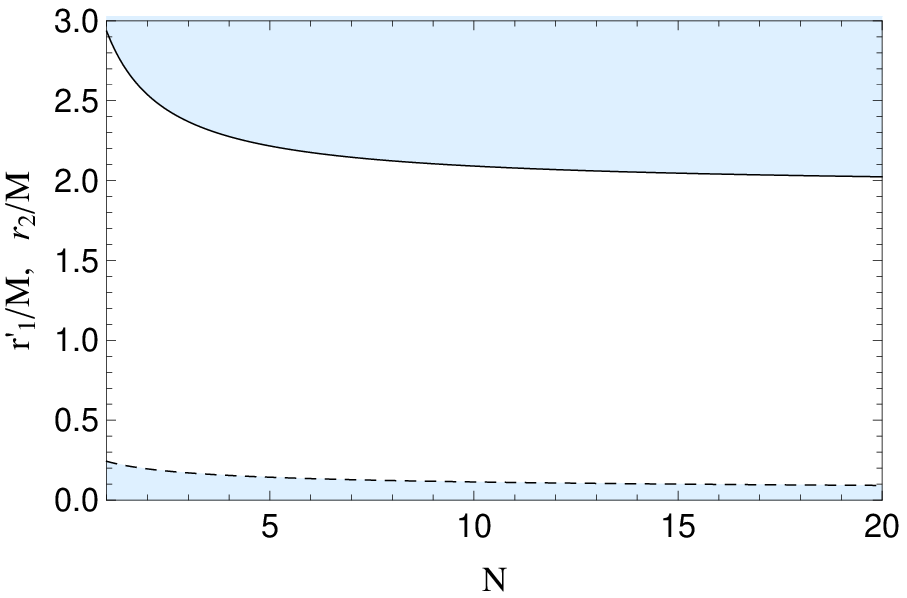}
\includegraphics[width=0.44\linewidth]{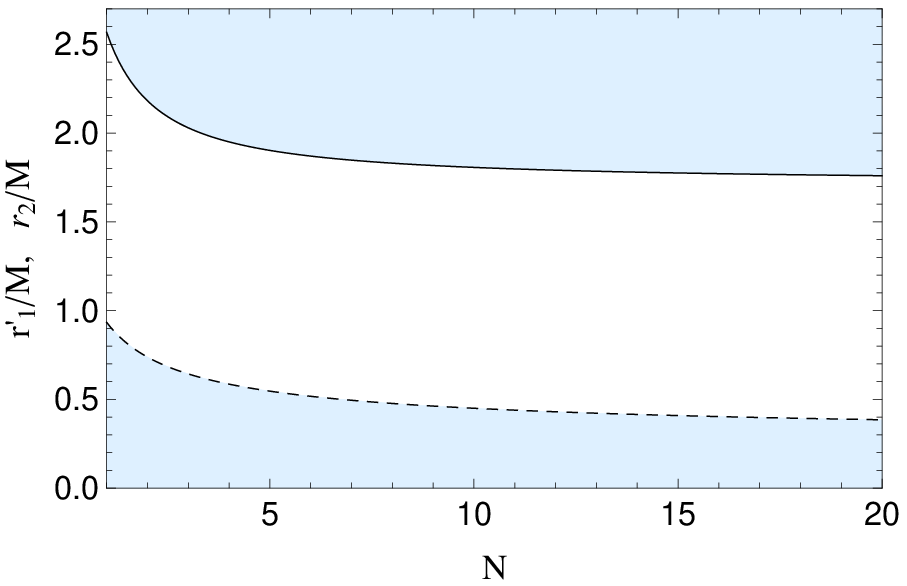}
\end{center}
\caption{\label{fig-nec-kerr2} Dependence of the boundaries of the NEC in the regular rotating black hole spacetime in conformal gravity on the degree $N$ in the large conformal factor limit. In the left panel we have $a/M=0.3$, while in the right panel $a/M=0.7$. Outer $r'_1$ (dashed curves) boundary of the inner branch and inner boundary $r_2$ (solid curves) of the outer branch of the NEC are depicted. In the shaded region the NEC is satisfied, while in the white region the NEC is violated.}
\end{figure*}

For large values of the parameter $L$, the relations~(\ref{nec-kerr-1}) are close to zero. However, still there are regions where the NEC is violated. In this case, the outer boundary of the outer branch of the NEC satisfied region, $r'_3$, tends to infinity, since $r'_3$ is a monotonically increasing function of $L$. However, other boundaries tend to a finite value depending on the values of the parameters $a$, $M$, and $N$. Due to the cumbersome form of the boundaries: $r'_1$ -- outer boundary of the inner branch of the NEC, $r_2$ -- inner boundary of the outer branch of the NEC, we present the results in Fig.~\ref{fig-nec-kerr2}.
The rotation of the spacetime decreases the depth of the violation of the NEC and if we increase the value of the rotation parameter, the NEC starts to be satisfied just inside and outside the Cauchy horizon.

Fig.~\ref{fig-nec-kerr2} shows the regions in which the NEC is violated and satisfied for large values of $N$. In the case of large values of $N$, like for the non-rotating case~(\ref{nec-schw-N}), the inner boundary of the outer branch of the NEC tends to the event horizon, while, the outer boundary of the inner branch of the NEC tends to the Cauchy horizon of the non-singular rotating black hole, i.e.,
\bear
&&r_2=M-\sqrt{M^2-a^2}+\mathcal{O}\left(\frac{1}{N}\right),\nonumber\\
&&r_3=M+\sqrt{M^2-a^2}+\mathcal{O}\left(\frac{1}{N}\right).
\ear
Thus, when the black hole is maximally rotating, $a=M$, the two branches of the NEC join.

\subsection{Weak Energy Condition}

The weak energy condition (WEC) can be written as
\bear
T_{\mu\nu}u^\mu u^\nu\geq0\ ,
\ear
where $u^\mu$ is a generic time-like vector. The physical interpretation of the WEC is that the total energy density of all matter fields measured by any observer traversing a timelike curve, is never negative. In the LNRF, it corresponds to
\bear\label{wec}
\rho\geq0, \qquad \rho+P_i\geq0.
\ear
One can see from~(\ref{nec}) and~(\ref{wec}) that the WEC requires not only that the NEC is satisfied, but also that the zeroth component of the energy-momentum tensor in the LNRF must be non-negative. Shortly,
\bear\label{wec-nec}
WEC\in[\rho\geq0\cap NEC].
\ear

\subsubsection{WEC for non-rotating black holes in conformal gravity}\label{wec-schwarzschild}

In the subsubsection~\ref{nec-schwarzschild} it has been shown that the NEC in the non-singular non-rotating black hole spacetime in conformal gravity is violated. From that fact we can assert that the WEC is also violated in this spacetime. In this subsubsection we aim at investigating the regions where the WEC is satisfied. Note that the WEC satisfied region is equal to the one of the NEC at most, otherwise smaller (see Eq.~(\ref{wec-nec})).

The 00-component of the energy-momentum tensor, $\rho$, is given in Appendix~\ref{A-schw1}. The non-negativity of $\rho$ gives the following inequality:
\bear\label{wec-schw-ineq}
r^2(r-3M)+L^2[M(1- 2N)+(N-1)r]\geq0\ ,
\ear
$\rho$ has always two non-vanishing zeros: $r^\ast_1=0$ and $r^\ast_2$. Combining $r^\ast_1=0$ with the fact that at the center of the spacetime the NEC is satisfied, we can assert that at the center of the spacetime the WEC is satisfied as well.

\begin{figure*}[th]
\begin{center}
\includegraphics[width=0.33\linewidth]{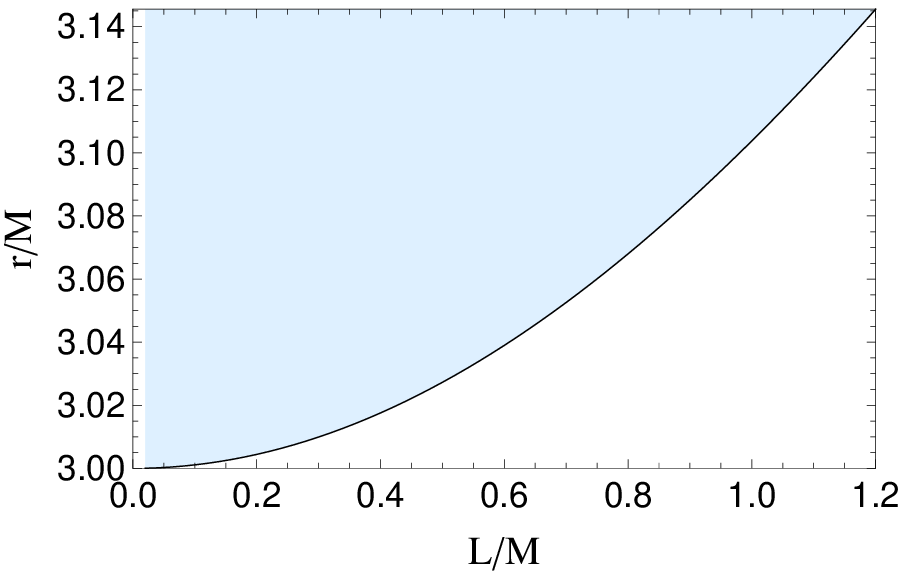}
\includegraphics[width=0.33\linewidth]{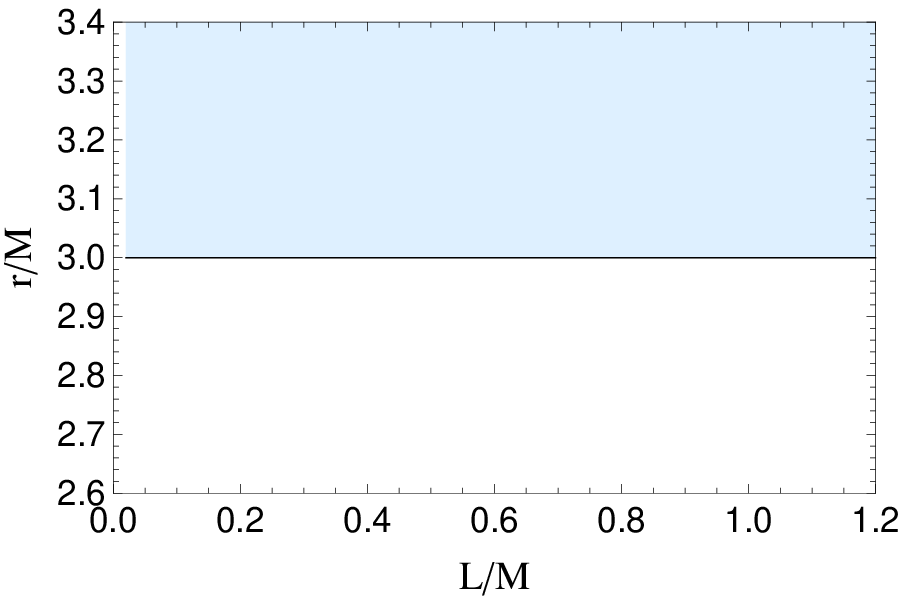}
\includegraphics[width=0.33\linewidth]{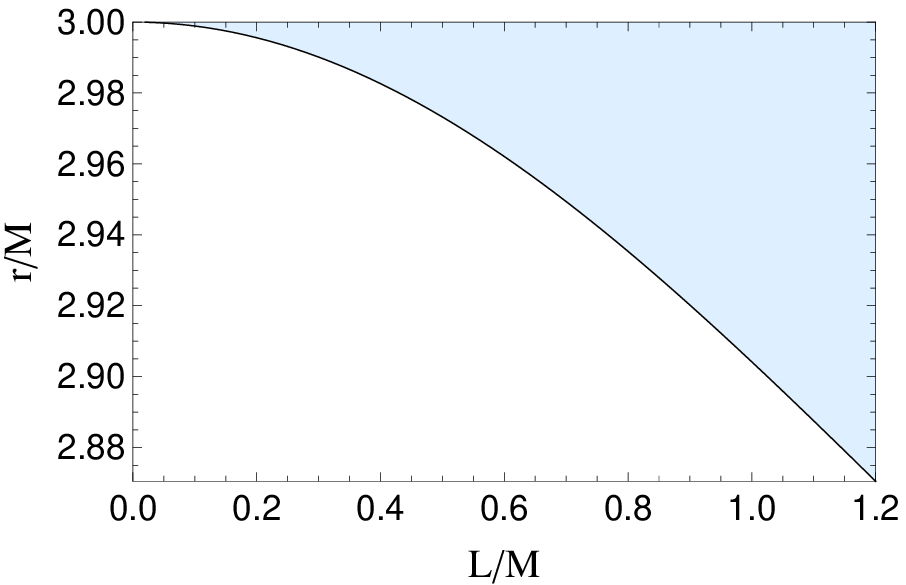}
\end{center}
\caption{\label{fig-wec-schw} Plot showing $\rho\geq0$ (shaded) region in the non-singular non-rotating black hole spacetime in conformal gravity. The boundary of the white and blue regions (solid black curve) represents $r^\ast_2$. Here the left, middle, and right panels are for $N=1$, $N=2$, and $N=3$, respectively.}
\end{figure*}

Moreover, at large distances, $\rho$ tends to zero with positive value
\bear
\lim_{r\rightarrow\infty}\rho=+0.
\ear
This indicates that $\rho\geq0$ for $r\in\{0\}\cup[r^\ast_2,+\infty)$.
Thus, the inequality $\rho\geq0$ does not determine the outer boundary for the WEC and the WEC is satisfied in the region $r\in\{0\}\cup[r'_2, r'_3 ]\cap[r_2, r'_3]\cap[r^\ast_2,r'_3]$ or $r\in\{0\}\cup[Max[r_2,r'_2,r^\ast_2], r'_3 ]$.
In Fig.~\ref{fig-wec-schw} the behaviour of $r^\ast_2$ ($\rho\geq0$) is depicted.
One can see from Fig.~\ref{fig-wec-schw} that in the case $N=1$, $r^\ast_2$ increases with $L/M$, however, in the case of $N=2$, $r^\ast_2$ does not depend on $L/M$ and does coincide with the radius of the unstable circular photon orbit, $r^\ast_2=3M$. For the cases $N>2$, $r^\ast_2$ decreases with $L/M$. As it was stated in subsubsection~\ref{nec-schwarzschild} concerning the constraint $L/M\leq1.2$, to have the NEC satisfied region except $r\in\{0\}$, $N$ must be $N\geq6$. Thus, this requires $N\geq6$ for the WEC too. For this values of $N$, $r^\ast_2>Max[r_2,r'_2]$, the WEC is satisfied in the region $r\in\{0\}\cup[r^\ast_2, r'_3 ]$.

\subsubsection{WEC for rotating black holes in conformal gravity}\label{wec-kerr}

In subsubsection~\ref{subsec-nec-kerr} the NEC satisfied regions in the non-singular rotating black hole spacetime in conformal gravity were studied. Now we do the same analysis for the 00-component of the energy-momentum tensor, $\rho$. The components of the energy-momentum tensor of the non-singular rotating black hole spacetime in conformal gravity are given in Appendix~\ref{A1}. At the center of the spacetime and spatial infinity, the 00-component of the energy-momentum tensor, $\rho$, tends to the positive finite value
\bear
&&\lim_{r\rightarrow0}\rho=\frac{4a^{4N+2}(N+1)L^2}{(a^2+L^2)^{2N+3}},\nonumber\\
&&\lim_{r\rightarrow\infty}\rho=+0,
\ear
$\rho$ has zero, one ($r^\ast_1=r^\ast_2$), and two ($r^\ast_1$, $r^\ast_2$) positive zeros depending on the values of the parameters $L$, $M$, $N$, and $a$. If it has one zero or does not have zero, $\rho\geq0$ condition is satisfied everywhere in the spacetime. This case corresponds to the rapidly rotating black hole, $0.8M\leq a\leq M$, or rotating no-horizon spacetimes, $a>M$. If $\rho$ has two zeros, the $\rho\geq0$ condition is satisfied in $r\in[0,r^\ast_1]\cup[r^\ast_2,+\infty)$. Thus, as in the non-rotating case, in the rotating case, $\rho\geq0$ does not put the upper boundary on the WEC. By combining the condition $\rho\geq0$ with the NEC we obtain the WEC satisfied region $r\in[0,r^\ast_1]\cap[0,r_1]\cap[0,r'_1]\cup[r_2,r'_3]\cap[r'_2,r'_3]\cap[r^\ast_2,r'_3]$.
Analysis have shown that $r^\ast_1>r_1,r'_1$ and $r^\ast_2<r_2,r'_2$. Therefore, the WEC in the rotating regular spacetime in conformal gravity is satisfied in the region where the NEC is satisfied.

\subsection{Strong Energy Condition}

The strong energy condition (SEC) implies
\bear
\left(T_{\mu\nu}-\frac{1}{2}Tg_{\mu\nu}\right)u^\mu u^\nu\geq0\ ,
\ear
In the LNRF, it assumes the form
\bear\label{sec}
\rho+\sum_{i=1}^3P_i\geq0.
\ear
Below we check the SEC in the non-singular non-rotating and rotating black hole spacetimes in conformal gravity.

\subsubsection{SEC for non-rotating black holes in conformal gravity}\label{sec-sec-schw}

\begin{figure}[h]
\centering
\includegraphics[width=0.43\textwidth]{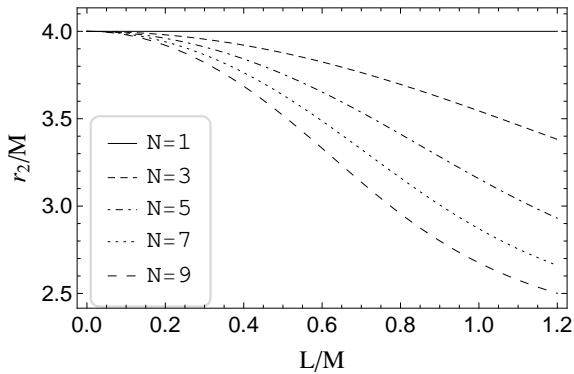}
\caption{\label{fig-sec-schw} Dependence of inner boundary of the SEC ($r_2$) in the non-singular non-rotating spacetime in conformal gravity on the parameters $L$ and $N$.}
\end{figure}

The sum of the components of energy-momentum tensor in the non-singular non-rotating black hole spacetime in conformal gravity is given by
\bear\label{sec-schw1}
\rho+\sum_{i=1}^3P_i&&=\frac{12NL^2r^{4N-3}}{(r^2+L^2)^{2N+2}} \\&&\times[r^2 (r-4M)-L^2(4MN+r-2Nr)],\nonumber
\ear
$\rho+\sum_{i=1}^3P_i$ has two zeros, $r_1=0$ and $r_2$, independently of the values of the parameters $L$ and $N$. $\rho+\sum_{i=1}^3P_i<0$ for $r\in(0,r_2)$ and $\rho+\sum_{i=1}^3P_i\geq0$ for $r\in\{0\}\cup[r_2,+\infty)$. We do not write here the full expression of $r_2$ because of its cumbersome form, but we show its dependence on the parameters $L$ and $N$ in Fig.~\ref{fig-sec-schw}.
The SEC is satisfied in the region $r\in\{0\}\cup[r_2,\infty)$. The size of the region increases for larger values of $L$ and $N$.

As we can see from Fig.~\ref{fig-sec-schw}, if we increase the values of $L$ and $N$, the inner boundary of the SEC region, $r_2$, decreases. For large values of $L$, it tends to the finite value
\bear
r_2=\frac{4MN}{2N-1}+\mathcal{O}\left(\frac{1}{L^2}\right)\ .
\ear
For large values of $N$, as in the case of the NEC and WEC, $r_2$ tends to the event horizon of the black hole
\bear
r_2=2M+\mathcal{O}\left(\frac{1}{N}\right)\ .
\ear
In the large $N$ limit, the only difference between the SEC region and the NEC/WEC region is that the outer boundary for the SEC region is not finite.

\subsubsection{SEC for rotating black holes in conformal gravity}\label{sec-sec-kerr}

\begin{figure*}[th]
\begin{center}
\includegraphics[width=0.44\linewidth]{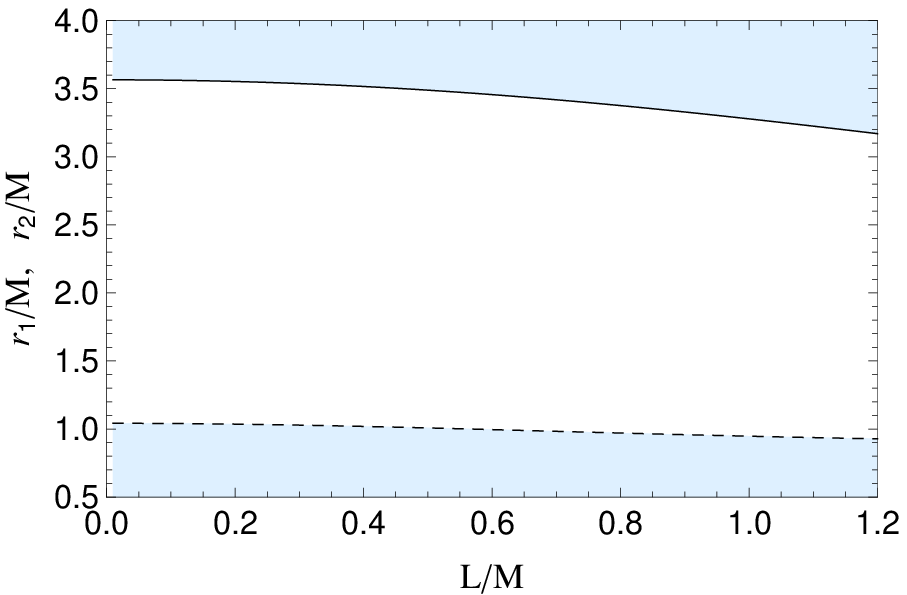}
\includegraphics[width=0.44\linewidth]{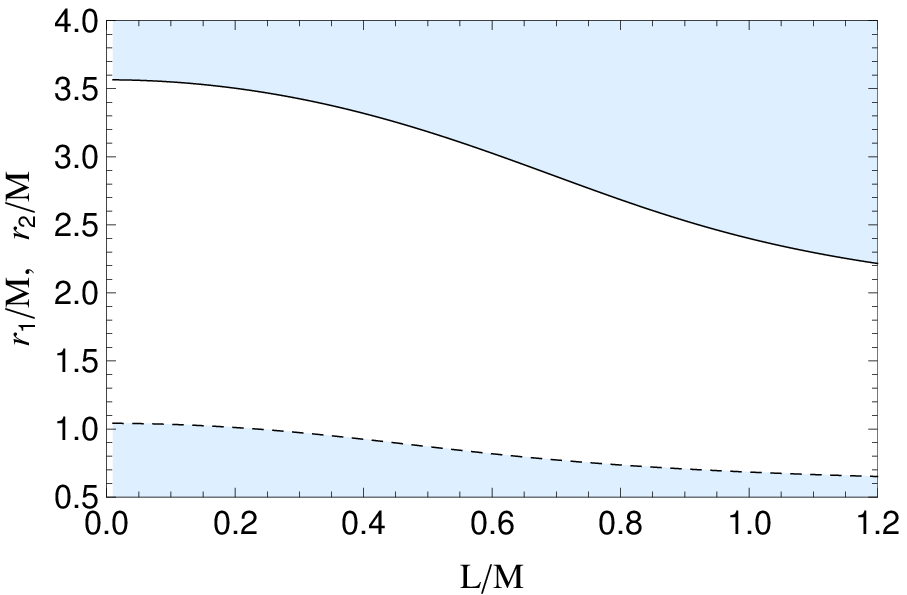}
\end{center}
\caption{\label{fig-sec-kerr1} Dependence of the boundaries of the SEC in the regular rotating black hole spacetime in conformal gravity on the dimensionless parameter $L/M$. In the left panel, we have $a/M=0.7$ and $N=1$, while in the right panel $a/M=0.7$ and $N=5$. The plots also show $r_1$ (dashed curve) and $r_2$ (solid curve). In the shaded region, the SEC is satisfied, while in the white region the SEC is violated.}
\end{figure*}

In the case of rotating black holes, $\rho+\sum_{i=1}^3P_i$ can have two zeros, $r_1$ and $r_2$. In the $r\rightarrow0$ limit, we find
\bear\label{sec-kerr-limit}
\lim_{r\rightarrow0}\left(\rho+\sum_{i=1}^3P_i\right)= \frac{12(N+1)a^{4N+2}L^2}{(a^2+L^2)^{2N+3}} \, ,
\ear
which is positive. $\rho+\sum_{i=1}^3P_i$ may have zero, one ($r_1=r_2$), or two ($r_1$, $r_2$) zeros. If it has no zero or one zero, $\rho+\sum_{i=1}^3P_i$ is positive everywhere in the spacetime, i.e., the SEC is satisfied everywhere in the spacetime and this case corresponds to the extremal rotating regular black hole spacetime, $a\approx M$, and the no-horizon spacetime, $a>M$. When $\rho+\sum_{i=1}^3P_i$ has two zeros, the SEC turns out to be satisfied in the region $r\in[0,r_1]\cup[r_2,+\infty)$. For the small values of the rotation parameter $r_1$ is close to zero and $r_2$ is close to $4M$ for $N=1$. With increasing value of the rotation parameter $r_1$ increases and $r_2$ decreases. In Fig.~\ref{fig-sec-kerr1}, the regions where the SEC is satisfied are depicted.
One can see from Fig.~\ref{fig-sec-kerr1} that increasing the values of the parameters $L$ and $N$, the values of the boundaries $r_1$ and $r_2$ decrease. Note that $r_2$ reaches its maximum, $r_2\rightarrow4M$, when the black hole is slowly rotating with small value of $L$.

\begin{figure*}[th]
\begin{center}
\includegraphics[width=0.44\linewidth]{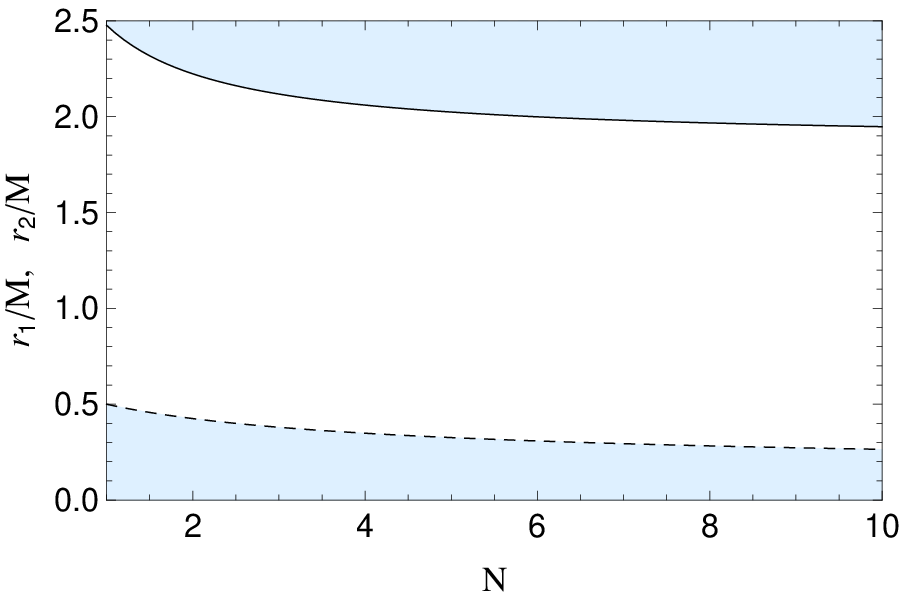}
\includegraphics[width=0.44\linewidth]{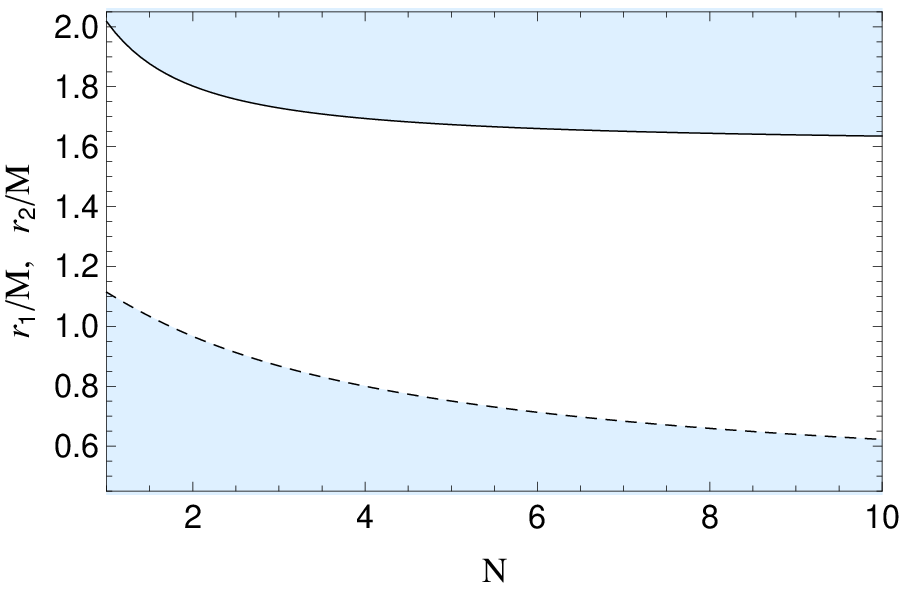}
\end{center}
\caption{\label{fig-sec-kerr2} Dependence of the boundaries of the SEC in the regular rotating black hole spacetime in conformal gravity on the dimensionless parameter $N$ in the large $L$ limit. Here we have $a/M=0.5$ (left panel) and $a/M=0.8$ (right panel). The plots also show $r_1$ (dashed curve) and $r_2$ (solid curve). In the shaded region the SEC is satisfied, while in the white region the SEC is violated.}
\end{figure*}

For large values of the parameter $L$, $r_1$ and $r_2$ tend to finite values. In Fig.~\ref{fig-sec-kerr2}, we show their behaviors in terms of the parameters $a$ and $N$.

For large values of $N$, $r_1$ and $r_2$ tend to the Cauchy and event horizons of the rotating black hole, respectively, as in the case of the NEC and WEC
\bear
&&r_1 = M - \sqrt{M^2-a^2} + O\left(\frac{1}{N}\right),\\
&&r_2 = M + \sqrt{M^2-a^2} + O\left(\frac{1}{N}\right).
\ear

If $\rho+\sum_{i=1}^3P_i$ has no zeros, the SEC is satisfied everywhere in the spacetime. This corresponds to the spacetimes without horizon ($a>M$). Thus, the rotation of the spacetime not only violates the energy conditions if they are satisfied when the spacetime is not rotating~\cite{Bambi:PLB:2013,Zaslavskii:PLB:2010,Neves:PLB:2014,Toshmatov:PRD:2014}, it increases the possibility of satisfaction of the energy conditions if they are violated when the spacetime is non-rotating.

\section{Conclusion}\label{conclusion}

In this paper we have studied the energy conditions of the non-rotating and rotating singularity-free black hole spacetimes found in Refs.~\cite{Bambi:1611.00865,Modesto:2016max}. More precisely, we have studied the energy conditions of the effective energy-momentum tensor generating this family of metrics assuming Einstein's equations.

In the presence of conformal symmetry, all the conformally-equivalent metrics describe the same physical spacetime. If conformal invariance is a fundamental symmetry of Nature, it must be broken, because the Universe is not conformally invariant. In the broken phase, some mechanism should select the true vacuum, and only one of the infinite solutions becomes the physical metric. However, we do not know currently how this can work. In this paper, we have studied the energy conditions of the effective energy-momentum tensor of these spacetimes to explore if we can use the energy conditions as a selection criterion to find the true vacuum, or at least to find the class of vacua the true vacuum may belong to. This is presumably the most general approach we can employ if we do not want to limit our discussion to a specific conformal gravity theory.

The NEC and the WEC are always violated near the center, near the (Cauchy and event) horizons, and at infinity. The SEC can be satisfied everywhere if the spacetime has no horizon, namely when $a \ge M$. Unfortunately, it does not seem to be possible to find a proper combination of $N$ and $L$ to satisfy some energy condition. Note that the vacuum solutions of Einstein gravity, namely the Schwarzschild and Kerr metrics, are recovered for $L=0$. However, these spacetimes are singular at the center. For a non-vanishing $L$, the spacetimes are regular everywhere, but the energy conditions are violated. This is true for any finite value of $L$.

The parameter $L$ is already constrained to be of order $M$ or less from astrophysical data of black hole candidates~\cite{Bambi:1701.00226}. With such a constraint, it seems that the ``best spacetimes'', namely those with minimum violation of the energy conditions, are the spacetimes with very large $N$. In these spacetimes, all the energy conditions are almost satisfied in the exterior region (outside of the black hole). While such a selection criterion based on the energy conditions of the effective energy-momentum tensor should be taken with great caution, our conclusion is that Nature may prefer a spacetime with a large $N$ as the true vacuum.

\section*{Acknowledgments}
B.T. and Z.S. would like to express their acknowledgments for the institutional support of the Faculty of Philosophy and Science of the Silesian University in Opava, the internal student grant of the Silesian University (Grant No. SGS/14/2016) and the Albert Einstein Centre for Gravitation and Astrophysics under the Czech Science Foundation (Grant No.~14-37086G). C.B. acknowledges support from the NSFC (grants 11305038 and U1531117), the Thousand Young Talents Program, and the Alexander von Humboldt Foundation. The research of B.A. and A.A. is supported in part by Grant No.~VA-FA-F-2-008 of the Uzbekistan Agency for Science and Technology, and by the Abdus Salam International Centre for Theoretical Physics through Grant No.~OEA-NT-01 and by the Volkswagen Stiftung, Grant No.~86 866. This research is partially supported by Erasmus+ exchange grant between SU and NUUz.

\appendix

\section{Energy-momentum tensor in the non-singular non-rotating black hole spacetime}

\bear\label{A-schw1}
\rho&=&\frac{NL^2r^{4N-3}}{2\pi(r^2+L^2)^{2N+2}}\nonumber\\
&&\times[r^2(r-3M)+L^2(M-2MN+(N-1)r)],\nonumber\\
P_1&=&\frac{NL^2r^{4N-3}}{2\pi(r^2+L^2)^{2N+2}}\\
&&\times\left[(3M-2r)r^2+L^2(3M(1-2N)+(3N-2)r)\right],\nonumber\\
P_2&=&\frac{NL^2r^{4N-3}}{2\pi(r^2+L^2)^{2N+2}}\nonumber\\
&&\times\left[2r^2(r-3M)+L^2(Nr-2M(N+1))\right]=P_3.\nonumber
\ear

\section{Energy-momentum tensor in the orthonormal tetrad frame (ZAMO)}\label{appendix1}

The expressions of the non-vanishing components of the energy-momentum tensor in the orthonormal tetrad (LNRF or Zero Angular Momentum Observer (ZAMO) frame)~(\ref{tetrad}) in the polar caps, $\theta=0, \pi$, is given by

\bear\label{A1}
\rho&=&\frac{(N+1)L^2(r^2+a^2)^{2N-1}}{2\pi(r^2+a^2+L^2)^{2N+4}}\{a^6+a^4[L^2+3r(M+ r)]\nonumber\\
&&+r^3[r^2(r-3M)-L^2 (M+2MN-Nr)]\nonumber\\
&&+a^2r[3r^3+L^2(3M+r+Nr)]\},\nonumber\\
P_1&=&\frac{(N+1)L^2(r^2+a^2)^{2N-1}}{2\pi(r^2+a^2+L^2)^{2N+4}}\{2 a^6+a^4[2L^2+r(M+2r)]\nonumber\\&&+r^3[(3M-2r)r^2+L^2(-3M-6MN+r+3Nr)]\\
&&+a^2r[2(2M-r)r^2+L^2(M+ 3(1+N)r)]\},\nonumber\\
P_2&=&\frac{(N+1)L^2(r^2+a^2)^{2N-1}}{2\pi(r^2+a^2+L^2)^{2N+4}}\{2 a^4(2M+r)\nonumber\\&&+a^2[-2(M-2 r)r^2+L^2(4M+r+Nr)]\nonumber\\&&+r^2[2r^2(r-3M)+L^2(r+Nr-2M(2+N))]\}=P_3.\nonumber
\ear

\label{lastpage}

\end{document}